# The Multimedia Project Quarked!

A. Bean, T. MacDonald
*University of Kansas, Lawrence, KS 66045, USA*

Can exposure to fundamental ideas about the nature of matter help motivate children in math and science and support the development of their understanding of these ideas later? Physicists, designers, and museum educators at the University of Kansas created the Quarked!™ Adventures in the subatomic Universe project to provide an opportunity for youth to explore the subatomic world in a fun and user friendly way. The project components include a website (www.quarked.org) and hands-on education programs. These are described and assessment results are presented. Questions addressed include the following. Can you engage elementary and middle school aged children with concepts related to particle physics? Can young children make sense of something they can't directly see? Do teachers think the material is relevant to their students?

## 1. What is Quarked?

Imagine a world in which kids were as excited about science as they are about their favorite animated characters on television and in movies. What would happen if instead of fictional characters, their animated heroes and nemeses were actually based on real scientific principles? Could we actually entertain kids while sparking their interest in science? A collaboration of physicists, educators, and designers at the University of Kansas originated the *Quarked!$^{TM}$ Adventures in the Subatomic Universe* project to explore this idea. The goal was to develop a multimedia project that would provide an engaging and educational science-based experience for youth aged 7 and up, educators and the general public. The project focuses on concepts of scale and matter, and presents subatomic particles as relatable characters in both human and quark or electron form that explore science through story-driven adventures. Figure 1 shows the high school kids with their quark counterparts.

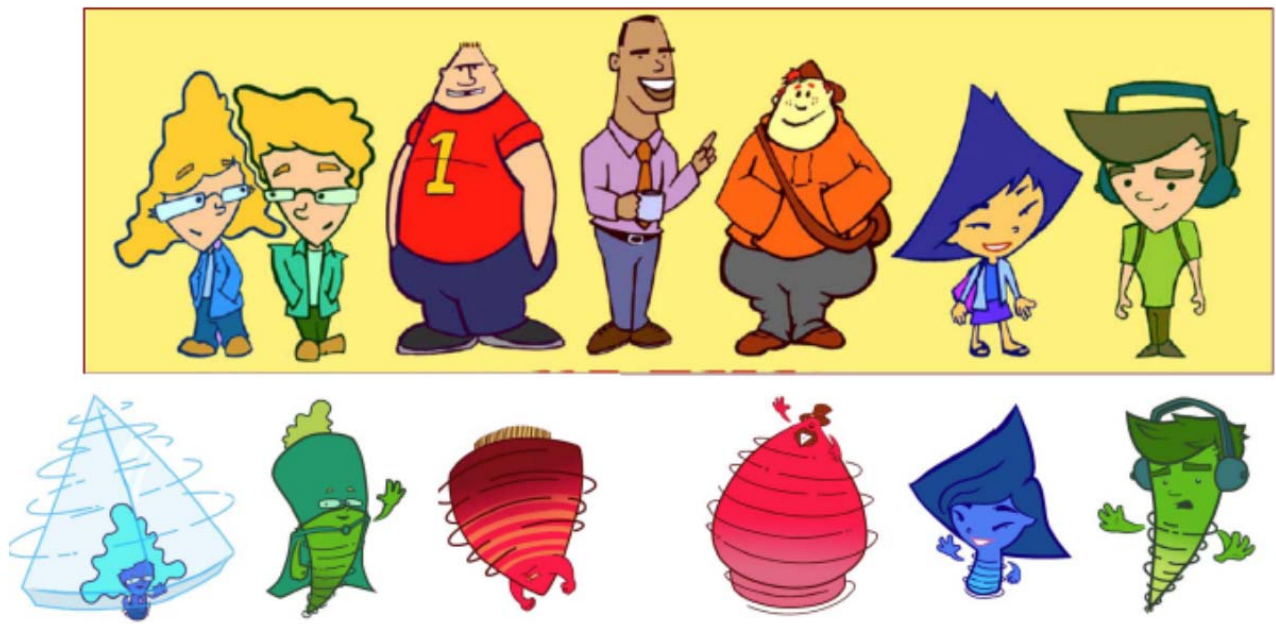

Figure 1. The main *Quarked!$^{TM}$* characters in their human form (on top) and as their quark alter ego (on bottom). In the middle of the top row is Mr. Marks who doesn't have a quark counterpart on the bottom. From left to right in the bottom row: Patti (charm quark), Matt (strange quark), Derek (bottom quark), Harold (top quark), Ushi (up quark), and Danny (down quark). Figure has Copyright 2006, the University of Kansas.

In 2005 the *Quarked!* team was awarded a National Science Foundation EPSCoR grant and matching support from the State of Kansas through the Kansas Technology Enterprise Corporation. The EPSCoR grant allowed the team to produce and launch an interactive web site with three 3D animated videos—including one that represents a preview for a proposed television series—design a series of online Flash games, create lesson plans and other online resources for teachers and parents, develop and present two hands-on science programs to thousands of 2nd to 8th grade students, as well as conduct formative and summative assessments on the material.

## 1.1 Website

The *Quarked!* interactive website [1] launched in 2006 and is continually evolving. The site introduces visitors to the world of subatomic physics and the Quarked! Project through:
- 3D video clips from the proposed television show
- *Game Zone* with activities for youth aged 7 and up, and the general public
- "Ask Mr. Marks" page filled with answers to many science questions, plus an opportunity for users to email their own questions to Quantum High School's cool and friendly science teacher
- Quarked! Club and Art Gallery with downloadable coloring pages and activities
- Educational outreach resources with lesson plans and other material for teachers and parents
- Links to information about current scientists and their research
- Online glossary,
- Information on the two Quarked! education programs available at the KU Natural History Museum

The website engages youth, parents, teachers, and others in a fun and educational science experience and serves as a central location for all of the *Quarked!* project components. A subsequent grant from the Ewing Marion Kauffman foundation expanded the website content at www.quarked.org. A Google$^{TM}$ Adwords grant has helped to promote and drive many people to the site. The *Quarked!* website has attracted over 100,000 unique visitors from over 60 countries. Public broadcasting television stations KTWU (Topeka, KS) and KCPT (Kansas City, MO) continue to broadcast the three animated videos available on the website as program interstitials.

## 1.2 Education programs

Since 2006, more than 5,000 youth have participated in the two hands-on museum programs—*How Small is Small* and *Quarks: Ups, Downs, and the Universe*). As part of a separate grant, a third program exploring how matter and forces are related to energy is currently under development, along with additional web resources for the *Quarked!* website. All programs incorporate demonstrations and hands-on activities and are taught by a professional museum educator.

In *How Small is Small*, we look at the idea of size and scale—how big we are compared to other things, and how small subatomic particles really are. We make comparisons to real-world objects and investigate the smallest things in the universe in multiple ways, including participants measuring their height in quarks. Also, why scale matters to how things work is discussed, such as why insects as tall as buildings could not exist.

*Quarks: Ups, Downs, and the Universe* introduces several key concepts including: particulate nature of matter, states of matter (solid, liquid, and gas), structure of atoms, the types of quarks and their properties—charge, mass and density. For example, in the density experiment, beakers are filled with three liquids of different densities to suspend six balls of the same size but different masses (representing the six quark flavors). This results in the heaviest or most dense on the bottom and the lightest or least dense floating on the top (see Fig. 2). Participants build atom models and explore the periodic table starting from ping-pong ball quarks and electrons.

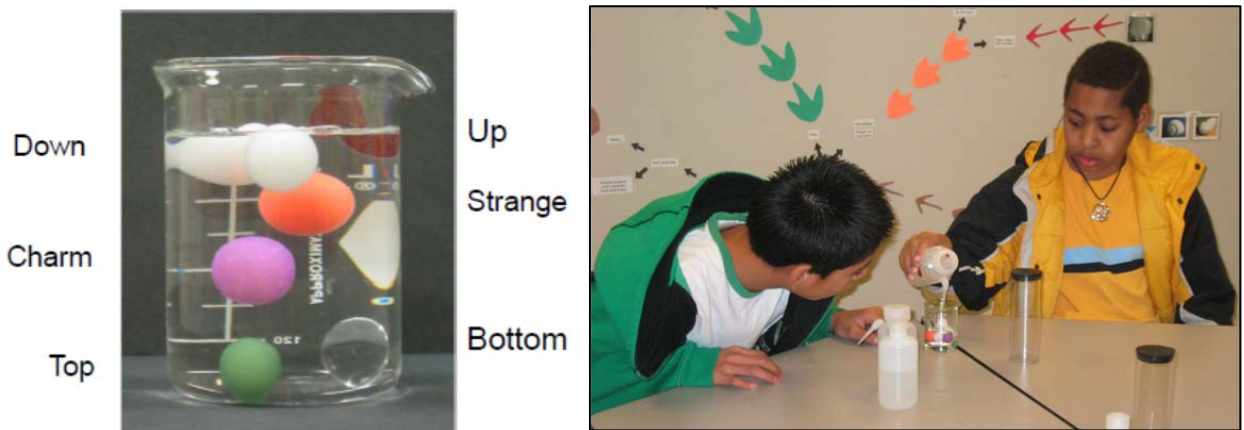

Figure 2. The left picture shows a beaker filled with different liquids (as shown in the right picture) to demonstrate density. There are six balls representing the six quarks.

## 2. Why Elementary Aged Children?

One may ask why this age group (7-12 years old) was chosen as the focus of this project. At this stage young people are open to everything and don't know that physics is "hard". Children are also curious and have a passion for learning, often interested in what they perceive to be unusual, unexpected or unique such as dinosaurs, distant planets and other far off objects in the sky. Early reviews of the project suggested that particle physics is too abstract and is more appropriately taught at the high school level. We didn't think this was correct. A literature review of relevant learning research was undertaken and a pilot assessment study was performed. The *Quarked!* team also promoted the project to K-12 teachers through workshops, regional conferences, and pedagogical publications [2].

Scale and the nature of matter are key scientific concepts that form the foundation for and support the understanding of other concepts, as well as current research across a range of sciences [3]. Futhermore, the learning research suggests that a greater emphasis early on in learning should be put on concepts of scale, and the nature and properties of matter and energy, such as the interactions of particles at and below the molecular level [4]. This research also suggests that concepts of small scale and the particulate nature of matter are accessible to young learners.

Our pilot study included [5] assessment of the animated videos and educational programs. One video assessment involved 7-11 year olds who were interviewed and observed, and another video assessment (ages 9-11) was conducted using a participant questionnaire. Educational programs were assessed through post-program teacher and student surveys (grades 2-8). The study focussed on three questions:
- Engagement—can one engage elementary and middle school aged children with concepts related to particle physics?
- Understanding—can young children make sense of something they can't directly see?
- Relevance—do teachers think the material is valuable for their students?

### 2.1 Assessment – Engagement

Can particle physics be engaging to elementary and middle school aged youth? Our study showed the resounding answer is yes.

Youth surveys showed that over 90% of the students rated the hands-on educational program as a 3 or higher (out of 5). More than 50% gave the highest rating of 5. When asked "What would you tell your friends about the class?", over half the students referred to sharing how much they enjoyed it (e.g. it was fun, cool or interesting), and 46% referred to some specific element of the program content that they would tell their friends about. One example response was "I had fun and they should have been there" (grade 5). In addition, video assessment showed that youth enjoyed their experience and were curious to find out more.

Student engagement through positive learning experiences that stimulate interest and enjoyment of science are critical to helping children seeing themselves as science learners, and can facilitate lifelong learning. We found that ideas related to particle physics can be made fun to elementary and middle school aged children if presented in an engaging and thoughtful way.

### 2.2 Assessment – Understanding

Can young children grasp the concept of something they can't directly see? We found the answer to be yes.

Video interviews and questionnaires demonstrate that children as young as 7 can grasp the idea of small things they cannot directly see, and that special equipment is needed. Education program surveys show many students' connection with the concepts explored with quotes such as: "Quarks are very, very tiny*"* (3rd grade); "I discovered that there is no possible way to make a giant bug that will take over the universe" (6th grade); "I learned that everything is made of matter and quarks are made of nothing else" (7th grade).

These findings support the idea that many young children can grasp the concept of the unseen, and are able to acknowledge and imagine the existence of quarks by the second grade.

## 2.3 Assessment – Relevance

Do teachers feel the ideas explored are relevant to their students and what they are currently doing? The teacher surveys showed us the answer was a definite yes.

All of the teachers felt the programs were valuable from both a curriculum and affective perspective and they all said that the programs provided a positive and fun science experience for their students. Some teacher comments collected: "Was wonderful; tied together some things I've already taught. Introduced new ideas and furthered knowledge." (2nd grade); "I think the kids were actively involved and also intrigued by new knowledge. They were also able to draw up on [lesson learned] just this year" (4th grade).

Our study has shown that relevant connections can be made between these science concepts and existing school curricula, and that teachers feel this material provides a positive learning experience for their students.

## 3. Summary

The *Quarked!*[TM] *Adventures in the Subatomic Universe* project was created at the University of Kansas to engage and excite youth about particle physics. Web resources have been created which include animated videos and games, and a range of other supporting material. The website continues to have over 5,000 unique visitors per month. New content is under development including a new *Photon Invaders* game exploring how solar panels work, a series of new simple animations about photons, electricity and superconductors, and material related to nanoscience. Hands-on education programs have reached more than 5,000 youth to date, and continue to be offered at the University of Kansas Museum of Natural History.

Teresa MacDonald and Alice Bean have published two papers based on results of assessments done with the project, which found that youth are engaged with the material and that teachers feel it supports student learning. Based on this pilot study, we advocate incorporating concepts related to particle physics earlier in education, and feel that it is important to create and support these kinds of enrichment experiences and opportunities for youth. Not only are such experiences valuable for younger students, but also finding ways to integrate science concepts that bridge the elementary through college curricula through informal learning experiences creates a great opportunity for lifelong learning.